# Prognostic classification based on random convolutional kernel

Zekun Wu[1] and Kaiwei Wu[2*]

1 Electrical & System Engineering, Washington University in St. Louis, St. Louis, USA; wu.zekun@wustl.edu
2 School of Port and Environmental Engineering, Jimei University, Xiamen, China; kaiwei_wu@outlook.com
* Correspondence: kaiwei_wu@outlook.com

**Abstract**
Assessing the health status (HS) of system/component has long been a challenging task in the prognostics and health management (PHM) study. Differed from other regression based prognostic task such as predicting the remaining useful life, the HS assessment is essentially a multi class classification problem. To address this issue, we introduced the random convolutional kernel-based approach, RandOm Convolutional KErnel Transforms (ROCKET) and its latest variant MiniROCKET, in the paper. We implement ROCKET and MiniROCKET on the NASA's CMAPSS dataset and assess the turbine fan engine's HS with the multi-sensor time-series data. Both methods show great accuracy when tackling the HS assessment task. More importantly, they demonstrate considerably efficiency especially compare with the deep learning-based method. We further reveal that the feature generated by random convolutional kernel can be combined with other classifiers such as support vector machine (SVM) and linear discriminant analysis (LDA). The newly constructed method maintains the high efficiency and outperform all the other deep neural network models in classification accuracy.

## 1. Introduction

Machine learning-based models have been widely used as prognostic approach to predictive maintenance in recent years [1], [2]. One of the most well studied areas is the remaining useful life (RUL) estimation, which benefited a lot from the great accuracy in machine learning method prediction [3]. Also, more complex machine learning models have proved to be useful in solving relevant topics such as anomaly detection and health status (HS) prognostic [4]–[6]. However, most of the previous studies are aimed at regression problem but lack of coverage on the prognostic classification, which could help solving lots of important industrial problems [7]. Applying machine learning-based classifier within the prognostic area cannot only solve the naturally defined classification problem, such as degradation assessment [8], but also deepen the understanding on the [7].

    Another general challenge when implementing the machine learning-based models is the slow training process that coming up with huge computational resources. In addition, the complicated structure within the model leads difficulty in choosing hyperparameter and requires heavy work in the hyperparameter tuning process. Evidently, there are a number of benefits in improving the efficiency of the machine learning models. One obvious one is that more efficient model brings faster testing speed on the hypothesis and thusly accelerating the process of scientific discoveries. Indeed, improving the machine learning-based models' efficiency become a relevant topic in the

artificial intelligence (AI) community in the last few years [9]. There are a few attempts to modify model structure, design novice algorithm or upgrade hardware to decrease the model training time in various application domains such as particle physics [10][11], biomedical engineering [12] and edge computing [13].

Among all the efforts to increase the machine learning methods' computation efficiency, one very promising approach for prognostic classification is the ROCKET (for RandOm Convolutional KErnel Transformer) [14]. The ROCKET is one of the newest classification model for solving the time series classification (TSC) problem. It achieves great accuracy on the dataset benchmark while maintaining high efficiency. The key idea behind ROCKET is to extract features from time-series with various random convolutional kernels. A concurrent work to ROCKET is the MiniROCKET [15], which restricted the hyperparameter space for convolution kernels. The MiniROCKET furtherly improves the processing time while having essentially the same accuracy.

In this work, we discuss prognostic classifiers based on the random convolutional kernel. We apply the ROCKET and MiniROCKET to assess the health status on the literature case study of CMAPSS turbofan engine [3], [16]. The results have been compared to those obtained by other machine learning methods. Also, we combine the random convolutional kernel generated features with different classifiers such as support vector machine (SVM) and linear discriminant analysis (LDA). The outcome of our study confirms that the random convolutional kernel-based classifiers have potential of prevailing prognostic analyses with great efficiency.

The remaining part of the paper is organized as follows: Sections 2 reviews related work to prognostic classification and efficient machine learning method. Section 3 describes the two general analytical phases of the random convolutional kernel-based classification method. Section 4 present the case study. Finally, conclusions and future work are addressed in Section 5.

## 2. Background

This section starts with reviewing the previous work in prognostic assessment. In session 2.1., we distinguish prognostic classification from the RUL prediction with introducing the health status (HS) assessment problem. Aiming at assessing the discrete HS of component/system, the approaches we discussed and implemented in the study are based exclusively on the classification methods from machine learning. Also, few attempts to increase the speed and efficiency of machine learning-based classification model are reviewed in session 2.2. Specifically, we focused on the latest developments in the multi time-series classification (TSC) study, which can help to reach good prognostic results while saving significant computation time.

### 2.1. Prognostic assessment: RUL prediction vs health status assessment

According to [3], the prognostic "refers to prediction/extrapolation/forecasting of process behavior, based on current health state assessment and future operating conditions". The exact definition of component/system health states is much dependent on the specific degradation process in the context. For example, the series work in lead-acid battery prognostic [17]–[20] choose state of charge (SOC), states of health (SOH) and states of function (SOF) as indicators to the battery's health states. Among all the relevant definitions of health states, the most associated concept is the RUL [21]. The realization of prognostics technique is related or even equal to the RUL estimation in various fields [22], [23]. A straightforward way to get the estimation result is inferring the RUL from the observed sensor data directly [24], [25]. One of the latest approach [25] learns the relationship between the selected sensors and the RUL of aero-engines by the extreme learning machine (ELM). Also, there are few attempts to improve the RUL estimation accuracy with constructing a health index (HI) [26]–[33]. In these methods, the HI is essentially an intermediate variable describing the component/system degradation pattern. It can be used as an input feature for the machine learning-based model [26], [29], [30]. For instance, Wu et al. extracts multiple degradation features from the rolling element bearings and use the principle component analysis (PCA) to reduce them into one HI variable [26]. The constructed HI is combined with both regression and Bayesian model to predict the bearings' RUL. When modelling the RUL, the state evolution is generally modelled as a continuous process. The RUL estimation, or HI-based RUL estimation, are all tackled as a regression problem in most cases.

The health status (HS) assessment is another crucial function in PHM [34]. Differed from the above-mentioned RUL estimation approach, the HS assessment targets on the discrete degradation process of component/system [32], [33], [35]–[37]. [32] proposes a subtractive-maximum entropy fuzzy clustering algorithm to identify the degradation states from unlabeled multi-dimensional data. Emmanuel et al. segments the input sequential data into four states and classifies the current system HS with a evidential Markoven classifier [35]. In [36], a HS classifier is built based on the stacked denoising autoencoder. The number of HS is determined by a grid search through the optimization of the classification results. In addition, the RUL and HS can be predicted in parallel way [8], [38], [39]. Zhang et al. construct a neural network model based on bidirectional gate recurrent unit (BIGRU) and multi-gate mixture-of-experts (MMoE) [39]. In the training process, the proposed model learns the predictable variables, the HS and the RUL, simultaneously. Experiment results show that the dual-task learning model not only reduces computational cost but also reaches good performance.

Noticeably, the HS estimation is intertwined with the RUL prediction [8]. The HS can play a very similar role to the HI and provide basis for next-stage's RUL prediction [32], [36]. To clarify what we are discuss here, the classification prognostic is defined as identifying the HS of component/system with supervised machine learning-based classification method. The HS are determined based on component/system's multiple degradation states. For the classification purpose, a multivariate label is assigned to each of the HS correspondingly. The problem is furtherly formulated in sessions 3.1.

## 2.2. Efficient machine learning model for prognostic classification

When applying the machine learning-based methods to solve real-world problem ,the use of complex model can be impractical as the requirement of huge computational resources [40][41]. To solve this issue, we would review few efficiency-oriented machine learning based approaches that can accelerate the training process in this session.

Proposed by Huang et al. [42], the ELM works on the feedforward neural networks in a relatively efficient way. It keeps the hidden layers untuned in the training process and improves the computational efficiency significantly [41]. The ELM has been proved capable of solving the prognostic problem. In [32][33], the summation wavelet-ELM (a ELM variant with dual activation functions in the hidden layer) is used to predict the RUL. The results show that ELM approach reaches a good balance between model accuracy and complexity. Echo state reservoir (ESN) is another promising method to increase the computation efficiency [43]. The ESN is based on the recurrent neural network (RNN) and preserves the RNN's capability in processing the time-series data. Similar to ELM, the hidden layers of ESN remains untrained during the adjustment process [40], [44]. In the series work for applying the RSS to RUL prediction [45], [46], it demonstrates great efficiency in computation.

As the classification prognostic problem is essentially a TSC problem, we also reviewed few state-of-art methods in this field. Unlike the domains such as computer vision, natural language processing, etc., the non-deep learning methods, which holds relatively less sophisticated structure and higher computational speed, can compete with the deep learning method in accuracy [47]. A representative one of them is the Hierarchical Vote Collective of Transformation-based Ensemble (HIVE-COTA). As an ensemble method , HIVE-COTA is firstly introduced in 2016 [48], [49]. It combines multiple classifiers and reaches significantly good accuracy. However, the great accuracy of HIVE-COTA is at the cost of huge computational resources, which let it infeasible to be implemented in certain cases. To accelerate the implementation speed of ensemble classifier, Shifaz et al. proposes a new TSC method called the Time Series Combination of Heterogeneous and Integrated Embedding Forest (TS-CHIEF) [50]. The TS-CHIEF uses a tree-based approach to speed up the training and testing process. The relevant results show that the TS-CHIEF competes HIVE-COTA in accuracy but requires only a fraction of the runtime. Among all the latest approach tackling the TSC problem, the most promising one to balance the accuracy and the efficiency is the ROCKET (for RandOm Convolutional KErnel Transform) [14]. According to [14], the ROCKET can be approximately hundred times faster than TS-CHIEF. More importantly, as far as we reviewed, ROCKET outperforms all the non-deep learning methods in accuracy when tackling the TSC problem. Even for the most advanced deep learning based model such as transformer, ROCKET can achieve a close performance while having advantage in learning speed [51].

## 3. Method
### 3.1. Problem formulation

We define the collected sensor data as multivariate time series: $X = \{x_1,…,x_m\} \in R^{m*l}$ with $x_i = (x_{i,1}, x_{i,2},…,x_{i,l})$, where $l$ is the time sequence length and $m$ is the multivariate dimensions number. We denote the $j$th timestamp of the $i$th time sequence of dimension $k$ as the scalar $x_{i,j,k}$. Following the definition in [8], [39], we assume the component/system's HS can be described by a discrete class variable $y$ with c possible values. Then, the classification prognostics is framed as a multivariate classification task in which the goal is to estimate the probability distribution over $y$ given X. It must be noted that the constructed HS label should be balanced among all the classes while reflecting the general degradation process. In this study, we follow the previous work to complete the HS label construction of the turbine fan engine [39].

### 3.2. Random convolutional kernel transform
Table 1 Hyperparameters comparisons between ROCKET and MiniROCKET

| Hyperparameter of kernel | ROCKET | MiniROCKET |
| --- | --- | --- |
| Length (l) | {7,9,11} | 9 |
| Weights (w) | N(0,1) | {-1,2} |
| Bias (b) | U(-1,1) | From convolution output |
| Dilation (d) | Random | Fixed |
| Padding (p) | Random | Fixed |

The implementation of ROCKET on time-series data can generally be divided into two phases. This session focused on the first phase, which starts with generating large numbers of random convolution kernels. These kernels are with random length, weights, bias, dilation, and padding. They convolve each time series and extract two important features from them: the maximum value

(MAX) and the proportion of positive values (PPV). For a given time series $x_i$ with $l$ time stamp, these two features can be defined in the following equations:

$$MAX = max\{x_i * c_i\}$$
$$PPV = \frac{1}{m}\sum_{i=1}^{m}[x_i * c_i + b > 0] \qquad (1)$$

where b is the bias scalar, [•] represents the Inversion bracket, $x_i$ is $i$th time sequence, $c_i$ is the random convolutional kernel used on the $i$th time sequence, and the $x_i * c_i = \sum_{j=1}^{l} x_{i,j} c_{i,j}$. The MAX is an equivalent to the global maximum pooling, while the PPV indicates the proportion of the input matching a given pattern. The PPV is the most critical element of ROCKET and contributes to its high accuracy.

The MiniROCKET is a reformulation of the ROCKET and can be operated in faster way. The major modification of MiniROCKET is that it fixes few hyperparameter of convolutional kernels. As indicated in the table 1, compared with the ROCKET, the MiniROCKET minimizes number of options for hyperparameters as length, weights, bias, dilation and padding. In addition, the MiniROCKET focus on the PPV without considering the MAX feature. As a result, it generates half number of features that ROCKET produces. Based on [15], the MiniROCKET is 75 faster than ROCKET.

### 3.3. High dimensional data classification methods

In the first phase, the ROCKET extracts feature with the random convolutional kernel. It turns the multi-variate time series classification into a general classification problem. Technically, the transformed features can be used with any classifier. In the previous studies, the ridge regression based classifier (for small dataset) and the stochastic regression classifier (for large dataset) are the most common choice for random convolution kernel transformed features [14], [15]. As a huge number of random convolution kernels (10,000 by default) are generated, the feature number is extremely large. This poses significant statistical challenges such as curse of dimensionality [52], poor generalization ability [53], etc. Actually, as stated in [54], the "impact of dimensionality on classification is largely poorly understood". In addition, to keep the main motivation of this study, we try to implement the classification model efficiently and do not consider any data preprocessing technique such as dimension reduction [55]. Thusly, we examine another two approaches that can tackle the high dimensional classification problem: the SVM and the LDA.

The SVM has been proved successful when making classification prediction on high dimensional datasets. The work of [56] shows that the SVM can reach high classification accuracy when being used with small training dataset but high-dimensional data. Similar good performance have been observed in the field of document classification and gene expression analysis [57], [58].

The LDA follows the Bayes rule to classify the input [59], [60]. It makes classification with maximizing the discrimination between the predefined groups. However, when tackling high dimensional data, the accuracy of LDA can be poor due to the singularity problem [61]. To solve this issue, one promising way is to introduce the shrinkage estimator to the LDA method [62]. In addition, the sparse LDA shows good results on high dimensional dataset [63].

### 4. Case study: HS assessment of aero-engine

This section will conduct a case study on applying the random convolutional kernel to a prognostic health management study. The dataset we are using is the NASA's CMAPSS data (commercial modular aero-propulsion system simulations) [64]–[66]. The goal of our experiments is to show the efficiency and accuracy of the random convolutional kernel transformer when estimating the HS of the turbine engine. Specifically, we conduct three prognostic classification experiments with different compared models and data samples on the CMPASS dataset. The first experiment (Exp 1) is on the FD001 sub dataset. In the experiment, we compare the performance of ROCKET and MiniROCKET with the most popular deep neural network model (RNN, GRU and LSTM) used in PHM. The second experiment (Exp 2) is referred to the work of Zhang et al. [39], in which the latest accuracy baselines for this HS classification problem have been created. In the Exp 2, we challenge these baseline on all four sub datasets. For the last experiment (Exp 3), we explore the kernel number's influence on the model's performance. Also, SVM and LDA have been combined with the transformed features to furtherly improve the model's accuracy.

### 4.1. Data preprocessing

Table 2 Overview of the turbofan datasets

| Sub Dataset | Operating condition | Fault mode |
|---|---|---|
| FD001 | 1 | 1 |
| FD002 | 6 | 1 |
| FD003 | 1 | 2 |
| FD004 | 6 | 2 |

As indicated in the Table 1, The CMPASS dataset contains four sub datasets (FD001, FD002, FD003 and FD004). As the simulated operating conditions and fault modes are differed in each sub dataset, the difficulty in making prediction gets increased from FD001 to FD004. The raw data provided in CMPASS are already in the form of multivariate time series, which records 21 sensors measurements on each turbine engine. Based on the previous work [67], [68], 14 sensor measurements out of the total 21 sensors are used as the input features. They are denoted as s2, s3, s4, s7, s8, s9, s11, s12, s13, s14, s15, s17, s20 and s21. In addition, the input data of each sensor is scaled with the Z-score standardization.

### 4.2. HS label construction

Following the work in [39], we constructed the health status label based on the following steps. Firstly, to produce the most relevant HI, the multi-dimensional sensory data are fused with the principle component analysis [69]. Then, the fused HI are smoothed with the S-G filter [70] and curved with the Weibull failure rate function [71]. Finally, the HS are divided based on the smoothed HI curve slope [39]. The visualized result of the HS label construction steps for FD001 are presented as figure below.

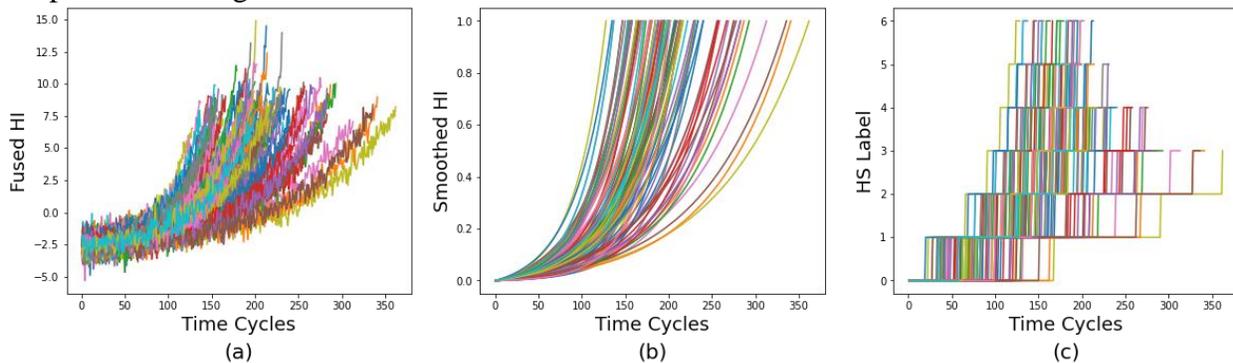

**Figure 1.** (a) Fused HI curve on FD001. (b) Smoothed HI curve. (c). HS labels of FD001.

### 4.3. Exp 1: Random convolution kernel model classification on FD001

Table 2. Model performance on FD001

| Model | ACC | $F1_{mac}$ | Computation time |
|---|---|---|---|
| LSTM | 34.57% | 20.92% | 858s |
| GRU | 29.87% | 17.35% | 694s |
| ROCKET | 46.26% | 41.13% | 23s |
| MiniROCKET | 47.56% | 38.17% | 5s |

In this session, we start with examining the performance of random convolution kernel model (ROCKET and MiniROCKET) on FD001. The kernel numbers for both models are set as 500.

To compare the random convolution kernel model with data-driven machine learning methods, we choose two deep neural network models: the GRU and the LSTM, as references in the experiment. The LSTM and the GRU are considered as the variants of the RNN. They are adept at processing dynamic information such as the time-series data [68]. The former one is preferred to prevent back-propagated errors from varnishing or exploding [72], while the latter one is with fewer parameters and therefore a faster learning process [73]. The LSTM and the GRU have proved to be valid in prognostic classification in the previous study [7], [39]. In the experiment, we keep both models sharing a similar structure of five stacked recurrent layers and one Softmax layer for multi classification.

To evaluate the experiment results, two evaluation metrics are used: the accuracy rate (ACC) and the Marco F1 ($F1_{mac}$). In addition, we keep the record of the average training time for all models. As indicated in the table, the ROCKET and MiniROCKET both reaches significantly better prediction results on the test set than the compared deep neural networks. However, what really makes difference here is the computing time. Based on the same hardware, the average training time of the LSTM model is 47s, while the ROCKET only takes 11s.

### 4.4. Exp 2: HS assessment on four sub datasets

To further examine the performance of the ROCKET on prognostic classification task, we conducted a comparative study with all the state-of-art methods used in [39]. It has to be noted that only training dataset are taken into account in the previous study [39]. We firstly split the FD001 training dataset into train and test set, then feed the ROCKET model (500 kernels) with all the newly created training samples. It takes ROCKET model 20 seconds to reach classification accuracy as 92.01%. By adjusting the number of kernels from 100 to 10000, the classification accuracy gets improved while the computation time getting increased as indicated in the Fig. 2. We eventually set the kernel number as 5000. Under The ROCKET model gets prediction results (ACC:93.77% $F1_{mac}$: 89.19%) with an average training time of 41s.

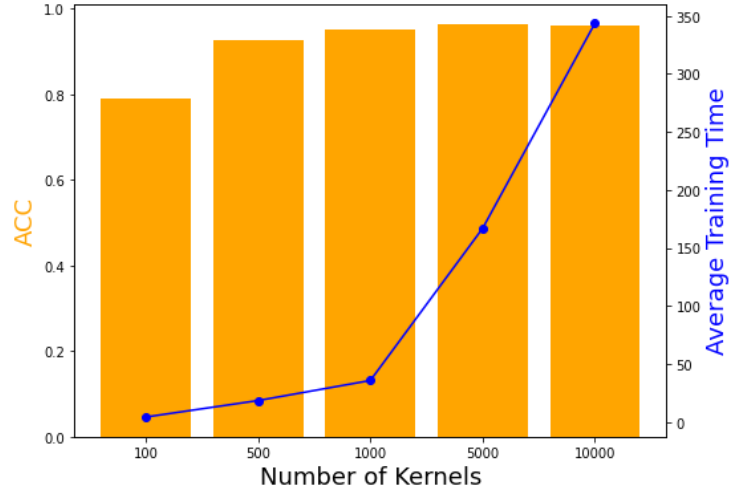

**Figure 2. Relationship between kernel numbers, ACC and average training time of ROCKET**

**Table 2. Model performance on resampled sub datasets**

| Model | FD001 | | FD002 | | FD003 | | FD004 | | CT |
|---|---|---|---|---|---|---|---|---|---|
| | ACC | F1$_{mac}$ | ACC | F1$_{mac}$ | ACC | F1$_{mac}$ | ACC | F1$_{mac}$ | |
| SB-BiGRU | 86.46% | 67.66% | 87.15% | 71.64% | 89.49% | 70.01% | **92.13%** | **80.32%** | 523s |
| MMoE-GRU | 77.17% | 57.21% | 68.25% | 47.65% | 77.58% | 55.74% | 78.12% | 60.55% | 461s |
| MMoE-BiGRU | 93.28% | 79.66% | 84.15% | 62.7% | 93.42% | 86.33% | 91.50% | 76.53% | 526s |
| ROCKET | **93.77%** | **89.19%** | **91.95%** | **83.12%** | **94.45%** | **89.45%** | 86.59% | 78.11% | 111s |
| Mini ROCKET | 91.76% | 83.57% | 86.86% | 79.54% | 90.02% | 85.25% | 74.04% | 66.55% | 25s |

Note: ACC: accuracy, CT: computation time

Then, we examine the ROCKET and MiniROCKET's performance on all the four sub datasets. As indicate in Table 2, compared with previous deep learning-based models (SB-BiGRU, MMoE-GRU and MMoE-BiGRU), both ROCKET and MiniROCKET show great effectiveness for tackling this HS assessment task. For sub datasets FD001, FD002 and FD003, the ROCKET outperforms all the other models with receiving the highest accuracy and F1$_{max}$ score. The MiniROCKET also achieves good performance on these three datasets but takes significantly less computation time. For FD004, the prediction accuracy of ROCKET and MiniROCKET both drop to some extent. The DNN (SB-BiGRU) method still holds the leading position on this dataset. As the FD004 is subjected to six operating conditions and two fault modes, which simulates the most complicated scenario in all the four sub datasets, the model performance decrease can be expected. Also, large number of extracted features bring extra challenge to the classification model. To further improve the classification results, we would test other classifiers capable of tackling high dimensional data on these four sub datasets in next session.

### 4.5. Exp3: High dimensional classification methods comparison

As mentioned in session 3.3, the number of extracted features with random convolutional kernels is very large. Based on the previous experiment results, we set the kernel number as 5000 and generates the exact same number of PPV features for HS classification. We implement two classification methods, the SVM and the LDA, for the same classification task in Exp 2. Compared with the Ridge classifier used in ROCKET and MiniROCKET, these two methods provide more accurate prediction results, as indicated in the Figure 3. We observe that both LDA and SVM achieve higher accuracy score than Ridge classifier on all four sub datasets. The SVM's $F1_{max}$ score on FD001 and FD003 falls behind the other two methods. Generally, LDA shows the best performance out of all three methods on this task. More importantly, the prediction results on the FD004 have been much improved with introducing the SVM (ACC: 96.25%, $F1_{max}$: 87.98%) and the LDA (ACC: 95.34%, $F1_{max}$: 90.70%).

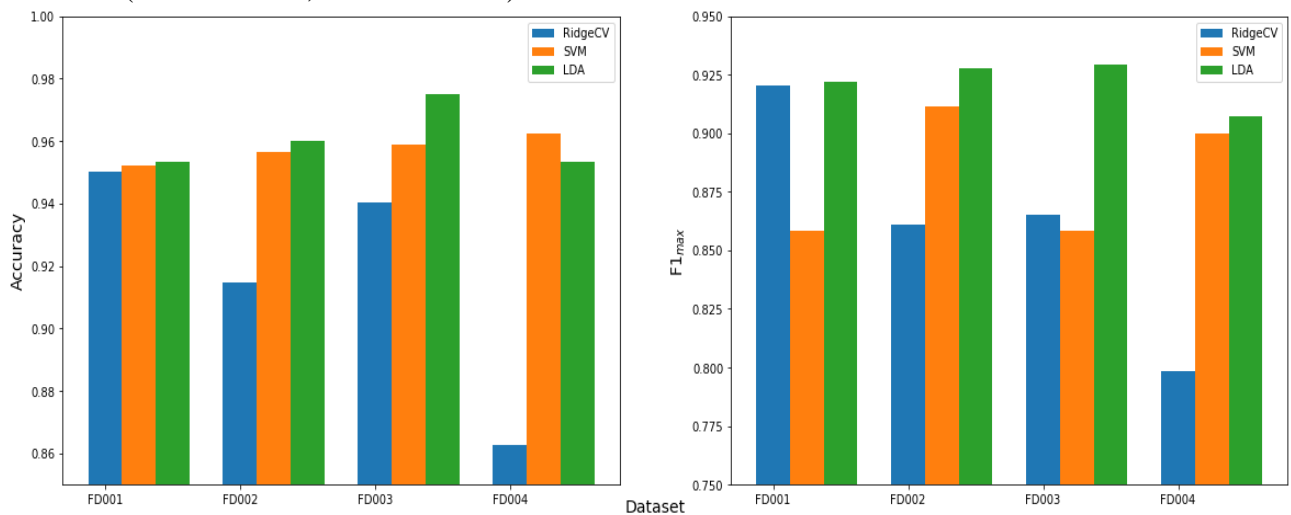

**Figure 3. Classification Performance of the Ridge classifier, SVM and LDA (Left: Accuracy, Right: $F1_{max}$)**

### 5. Conclusion and Future Work

In this work, we have implemented the random convolutional kernel-based approach, the ROCKET and MiniROCKET, on the HS assessment task. We show that both ROCKET and MiniROCKET are capable to tackle the prognostic classification problem with great efficiency. In the case study concerning the HS assessment, the computation time of random convolutional kernel-based method is significantly shorter than the DNN based model. Also, the random convolutional kernel transformed feature (PPV) is proved to be able to capture important information within the time series. These features can be combined with the methods that are generally being used in high-dimensional classification task. We reveal that applying the SVM and LDA on the random convolutional kernel feature space can furtherly improve the classification accuracy. As a part of future work, we would like to construct a two-stages pipeline method including the random convolutional kernel generation and the classification model selection. The idea is to search for the optimal hyperparameters in both stages and produce more accurate classification results, which can facilitate the current prognostic classification task.

[1]  X. Li, W. Zhang, and Q. Ding, "Deep learning-based remaining useful life estimation of bearings using multi-scale feature extraction," *Reliab. Eng. Syst. Saf.*, vol. 182, no. October 2018, pp. 208–218, 2019, doi: 10.1016/j.ress.2018.11.011.

[2]  Z. Xu and J. H. Saleh, "Machine Learning for Reliability Engineering and Safety Applications : Review of Current Status and Future Opportunities," no. Ml, pp. 1–50.

[3]  K. Javed, R. Gouriveau, and N. Zerhouni, "State of the art and taxonomy of prognostics approaches, trends of prognostics applications and open issues towards maturity at different technology readiness levels," *Mech. Syst. Signal Process.*, vol. 94, pp. 214–236, 2017, doi: 10.1016/j.ymssp.2017.01.050.

[4]  T. Brotherton and T. Johnson, "Anomaly detection for advanced military aircraft using neural networks," *IEEE Aerosp. Conf. Proc.*, vol. 6, pp. 63113–63123, 2001, doi: 10.1109/aero.2001.931329.

[5]  A. Listou Ellefsen, P. Han, X. Cheng, F. T. Holmeset, V. Aesoy, and H. Zhang, "Online Fault Detection in Autonomous Ferries: Using Fault-Type Independent Spectral Anomaly Detection," *IEEE Trans. Instrum. Meas.*, vol. 69, no. 10, pp. 8216–8225, 2020, doi: 10.1109/TIM.2020.2994012.

[6]  W. Yu, I. Y. Kim, and C. Mechefske, "An improved similarity-based prognostic algorithm for RUL estimation using an RNN autoencoder scheme," *Reliab. Eng. Syst. Saf.*, vol. 199, no. February, p. 106926, 2020, doi: 10.1016/j.ress.2020.106926.

[7]  M. L. Baptista, E. M. P. Henriques, and H. Prendinger, "Classification prognostics approaches in aviation," *Meas. J. Int. Meas. Confed.*, vol. 182, no. May, p. 109756, 2021, doi: 10.1016/j.measurement.2021.109756.

[8]  H. Miao, B. Li, C. Sun, and J. Liu, "Joint Learning of Degradation Assessment and RUL Prediction for Aeroengines via Dual-Task Deep LSTM Networks," *IEEE Trans. Ind. Informatics*, vol. 15, no. 9, pp. 5023–5032, 2019, doi: 10.1109/tii.2019.2900295.

[9]  A. M. Deiana *et al.*, "Applications and Techniques for Fast Machine Learning in Science," vol. 1, pp. 1–101, 2021, [Online]. Available: http://arxiv.org/abs/2110.13041.

[10] C. N. Coelho *et al.*, "Automatic heterogeneous quantization of deep neural networks for low-latency inference on the edge for particle detectors," *Nat. Mach. Intell.*, vol. 3, no. 8, pp. 675–686, 2021, doi: 10.1038/s42256-021-00356-5.

[11] P. Bedaque *et al.*, "A.I. for nuclear physics," *Eur. Phys. J. A*, vol. 57, no. 3, pp. 1–27, 2021, doi: 10.1140/epja/s10050-020-00290-x.

[12] Y. Zhang, G. Chen, H. Du, X. Yuan, M. Cheriet, and M. Kadoch, "Real-time remote health monitoring system driven by 5G MEC-IOT," *Electron.*, vol. 9, no. 11, pp. 1–17, 2020, doi: 10.3390/electronics9111753.

[13] Y. Sinan Nasir and D. Guo, "Deep Actor-Critic Learning for Distributed Power Control in Wireless Mobile Networks," *Conf. Rec. - Asilomar Conf. Signals, Syst. Comput.*, vol. 2020-Novem, pp. 398–402, 2020, doi: 10.1109/IEEECONF51394.2020.9443301.

[14] A. Dempster, F. Petitjean, and G. I. Webb, "ROCKET: exceptionally fast and accurate time series classification using random convolutional kernels," *Data Min. Knowl. Discov.*, vol. 34, no. 5, pp. 1454–1495, 2020, doi: 10.1007/s10618-020-00701-z.

[15] A. Dempster, D. F. Schmidt, and G. I. Webb, "MiniRocket: A Very Fast (Almost) Deterministic Transform for Time Series Classification," *Proc. ACM SIGKDD Int. Conf. Knowl. Discov. Data Min.*, pp. 248–257, 2021, doi: 10.1145/3447548.3467231.

[16] S. Uckun, K. Goebel, and P. J. F. Lucas, "Standardizing research methods for

prognostics," *2008 Int. Conf. Progn. Heal. Manag. PHM 2008*, 2008, doi: 10.1109/PHM.2008.4711437.

[17] I. S. Kim, "A technique for estimating the state of health of lithium batteries through a dual-sliding-mode observer," *IEEE Trans. Power Electron.*, vol. 25, no. 4, pp. 1013–1022, 2010, doi: 10.1109/TPEL.2009.2034966.

[18] L. Lu, X. Han, J. Li, J. Hua, and M. Ouyang, "A review on the key issues for lithium-ion battery management in electric vehicles," *J. Power Sources*, vol. 226, pp. 272–288, 2013, doi: 10.1016/j.jpowsour.2012.10.060.

[19] P. Shen, M. Ouyang, L. Lu, J. Li, and X. Feng, "The co-estimation of state of charge, state of health, and state of function for lithium-ion batteries in electric vehicles," *IEEE Trans. Veh. Technol.*, vol. 67, no. 1, pp. 92–103, 2018, doi: 10.1109/TVT.2017.2751613.

[20] H. Meng and Y. F. Li, "A review on prognostics and health management (PHM) methods of lithium-ion batteries," *Renew. Sustain. Energy Rev.*, vol. 116, no. January, p. 109405, 2019, doi: 10.1016/j.rser.2019.109405.

[21] X. S. Si, W. Wang, C. H. Hu, and D. H. Zhou, "Remaining useful life estimation - A review on the statistical data driven approaches," *Eur. J. Oper. Res.*, vol. 213, no. 1, pp. 1–14, 2011, doi: 10.1016/j.ejor.2010.11.018.

[22] Z. S. Chen, Y. M. Yang, and Z. Hu, "A technical framework and roadmap of embedded diagnostics and prognostics for complex mechanical systems in prognostics and health management Ssystems," *IEEE Trans. Reliab.*, vol. 61, no. 2, pp. 314–322, 2012, doi: 10.1109/TR.2012.2196171.

[23] C. S. Kulkarni, J. R. Celaya, G. Biswas, and K. Goebel, "Physics based degradation modeling and prognostics of electrolytic capacitors under electrical overstress conditions," *AIAA Infotech Aerosp. (I A) Conf.*, pp. 1–17, 2013, doi: 10.2514/6.2013-5137.

[24] X. Li, Q. Ding, and J. Q. Sun, "Remaining useful life estimation in prognostics using deep convolution neural networks," *Reliab. Eng. Syst. Saf.*, vol. 172, pp. 1–11, 2018, doi: 10.1016/j.ress.2017.11.021.

[25] J. M. Bai, G. S. Zhao, and H. J. Rong, "Novel direct remaining useful life estimation of aero-engines with randomly assigned hidden nodes," *Neural Comput. Appl.*, vol. 32, no. 18, pp. 14347–14358, 2020, doi: 10.1007/s00521-019-04478-1.

[26] J. Wu, C. Wu, S. Cao, S. W. Or, C. Deng, and X. Shao, "Degradation Data-Driven Time-To-Failure Prognostics Approach for Rolling Element Bearings in Electrical Machines," *IEEE Trans. Ind. Electron.*, vol. 66, no. 1, pp. 529–539, 2019, doi: 10.1109/TIE.2018.2811366.

[27] Y. Wei, D. Wu, and J. Terpenny, "Learning the health index of complex systems using dynamic conditional variational autoencoders," *Reliab. Eng. Syst. Saf.*, vol. 216, no. December 2020, p. 108004, 2021, doi: 10.1016/j.ress.2021.108004.

[28] X. Jia, C. Jin, M. Buzza, W. Wang, and J. Lee, "Wind turbine performance degradation assessment based on a novel similarity metric for machine performance curves," *Renew. Energy*, vol. 99, no. December 2017, pp. 1191–1201, 2016, doi: 10.1016/j.renene.2016.08.018.

[29] M. M. Manjurul Islam, A. E. Prosvirin, and J. M. Kim, "Data-driven prognostic scheme for rolling-element bearings using a new health index and variants of least-square support vector machines," *Mech. Syst. Signal Process.*, vol. 160, no. 7, p. 107853, 2021, doi: 10.1016/j.ymssp.2021.107853.

[30] F. Yang, M. S. Habibullah, and Y. Shen, "Remaining useful life prediction of induction

motors using nonlinear degradation of health index," *Mech. Syst. Signal Process.*, vol. 148, p. 107183, 2021, doi: 10.1016/j.ymssp.2020.107183.

[31] T. Yan, D. Wang, M. Zheng, T. Xia, E. Pan, and L. Xi, "Fisher's discriminant ratio based health indicator for locating informative frequency bands for machine performance degradation assessment," *Mech. Syst. Signal Process.*, vol. 162, no. December 2020, p. 108053, 2021, doi: 10.1016/j.ymssp.2021.108053.

[32] K. Javed, R. Gouriveau, and N. Zerhouni, "Novel failure prognostics approach with dynamic thresholds for machine degradation," *IECON Proc. (Industrial Electron. Conf.*, pp. 4404–4409, 2013, doi: 10.1109/IECON.2013.6699844.

[33] K. Javed, R. Gouriveau, and N. Zerhouni, "A new multivariate approach for prognostics based on extreme learning machine and fuzzy clustering," *IEEE Trans. Cybern.*, vol. 45, no. 12, pp. 2626–2639, 2015, doi: 10.1109/TCYB.2014.2378056.

[34] R. Li, W. J. C. Verhagen, and R. Curran, "A systematic methodology for Prognostic and Health Management system architecture definition," *Reliab. Eng. Syst. Saf.*, vol. 193, no. December 2018, p. 106598, 2020, doi: 10.1016/j.ress.2019.106598.

[35] E. Ramasso and R. Gouriveau, "Prognostics in switching systems: Evidential Markovian classification of real-time neuro-fuzzy predictions," *2010 Progn. Syst. Heal. Manag. Conf. PHM '10*, 2010, doi: 10.1109/PHM.2010.5413442.

[36] M. Xia, T. Li, T. Shu, J. Wan, C. W. De Silva, and Z. Wang, "A Two-Stage Approach for the Remaining Useful Life Prediction of Bearings Using Deep Neural Networks," *IEEE Trans. Ind. Informatics*, vol. 15, no. 6, pp. 3703–3711, 2019, doi: 10.1109/TII.2018.2868687.

[37] R.-J. Bao, H.-J. Rong, Z.-X. Yang, and B. Chen, "A Novel Prognostic Approach for RUL Estimation With Evolving Joint Prediction of Continuous and Discrete States," *IEEE Trans. Ind. Informatics*, vol. 15, no. 9, pp. 5089–5098, 2019, doi: 10.1109/tii.2019.2896288.

[38] E. Ramasso, M. Rombaut, and N. Zerhouni, "Joint prediction of continuous and discrete states in time-series based on belief functions," *IEEE Trans. Cybern.*, vol. 43, no. 1, pp. 37–50, 2013, doi: 10.1109/TSMCB.2012.2198882.

[39] Y. Zhang, Y. Xin, Z. Liu, M. Chi, and G. Ma, "Health status assessment and remaining useful life prediction of aero-engine based on BiGRU and MMoE," *Reliab. Eng. Syst. Saf.*, vol. 220, no. November 2021, p. 108263, 2021, doi: 10.1016/j.ress.2021.108263.

[40] M. Lukoševičius and H. Jaeger, "Reservoir computing approaches to recurrent neural network training," *Comput. Sci. Rev.*, vol. 3, no. 3, pp. 127–149, 2009, doi: 10.1016/j.cosrev.2009.03.005.

[41] G. Bin Huang, D. H. Wang, and Y. Lan, "Extreme learning machines: A survey," *Int. J. Mach. Learn. Cybern.*, vol. 2, no. 2, pp. 107–122, 2011, doi: 10.1007/s13042-011-0019-y.

[42] G. Bin Huang, Q. Y. Zhu, and C. K. Siew, "Extreme learning machine: A new learning scheme of feedforward neural networks," *IEEE Int. Conf. Neural Networks - Conf. Proc.*, vol. 2, pp. 985–990, 2004, doi: 10.1109/IJCNN.2004.1380068.

[43] H. Jaeger, "The" echo state" approach to analysing and training recurrent neural networks-with an erratum note'," *Bonn, Ger. Ger. Natl. Res. Cent. Inf. Technol. GMD Tech. Rep.*, vol. 148, 2001.

[44] H. Siqueira, L. Boccato, R. Attux, and C. Lyra, "Echo State Networks and Extreme Learning Machines: A Comparative Study on Seasonal Streamflow Series Prediction," in *Neural Information Processing*, 2012, pp. 491–500.

[45] M. Rigamonti, P. Baraldi, E. Zio, I. Roychoudhury, K. Goebel, and S. Poll, "Echo State Network for the Remaining Useful Life Prediction of a Turbofan Engine," *Annu. Conf. Progn. Heal. Manag. Soc.*, pp. 255–270, 2016.

[46] M. Rigamonti, P. Baraldi, E. Zio, I. Roychoudhury, K. Goebel, and S. Poll, "Ensemble of optimized echo state networks for remaining useful life prediction," *Neurocomputing*, vol. 281, pp. 121–138, 2018, doi: 10.1016/j.neucom.2017.11.062.

[47] A. Bagnall, J. Lines, A. Bostrom, J. Large, and E. Keogh, "The great time series classification bake off: a review and experimental evaluation of recent algorithmic advances," *Data Min. Knowl. Discov.*, vol. 31, no. 3, pp. 606–660, 2017, doi: 10.1007/s10618-016-0483-9.

[48] J. Lines, S. Taylor, and A. Bagnall, "Time series classification with HIVE-COTE: The hierarchical vote collective of transformation-based ensembles," *ACM Trans. Knowl. Discov. Data*, vol. 12, no. 5, 2018, doi: 10.1145/3182382.

[49] J. Lines, S. Taylor, and A. Bagnall, "HIVE-COTE: The hierarchical vote collective of transformation-based ensembles for time series classification," *Proc. - IEEE Int. Conf. Data Mining, ICDM*, pp. 1041–1046, 2017, doi: 10.1109/ICDM.2016.74.

[50] A. Shifaz, C. Pelletier, F. Petitjean, and G. I. Webb, "TS-CHIEF: a scalable and accurate forest algorithm for time series classification," *Data Min. Knowl. Discov.*, vol. 34, no. 3, pp. 742–775, 2020, doi: 10.1007/s10618-020-00679-8.

[51] G. Zerveas, S. Jayaraman, D. Patel, A. Bhamidipaty, and C. Eickhoff, *A Transformer-based Framework for Multivariate Time Series Representation Learning*, vol. 1, no. 1. Association for Computing Machinery, 2021.

[52] N. P. Patel, E. Sarraf, and M. H. Tsai, "The Curse of Dimensionality," *Anesthesiology*, vol. 129, no. 3, pp. 614–615, 2018, doi: 10.1097/ALN.0000000000002350.

[53] R. Clarke *et al.*, "The properties of high-dimensional data spaces: implications for exploring gene and protein expression data," *Nat. Rev. Cancer*, vol. 8, no. 1, pp. 37–49, 2008, doi: 10.1038/nrc2294.

[54] J. Fan, Y. Fan, and Y. Wu, "High-Dimensional Classification," pp. 3–37, 2010, doi: 10.1142/9789814324861_0001.

[55] S. Chowdhury, X. Dong, and X. Li, "Recurrent Neural Network Based Feature Selection for High Dimensional and Low Sample Size Micro-array Data," *Proc. - 2019 IEEE Int. Conf. Big Data, Big Data 2019*, pp. 4823–4828, 2019, doi: 10.1109/BigData47090.2019.9006432.

[56] M. Pal and P. M. Mather, "Support vector machines for classification in remote sensing," *Int. J. Remote Sens.*, vol. 26, no. 5, pp. 1007–1011, 2005, doi: 10.1080/01431160512331314083.

[57] M. Marseguerra, "Early detection of gradual concept drifts by text categorization and Support Vector Machine techniques: The TRIO algorithm," *Reliab. Eng. Syst. Saf.*, vol. 129, pp. 1–9, 2014, doi: 10.1016/j.ress.2014.03.014.

[58] S. Ramaswamy *et al.*, "Multiclass cancer diagnosis using tumor gene expression signatures," *Proc. Natl. Acad. Sci. U. S. A.*, vol. 98, no. 26, pp. 15149–15154, 2001, doi: 10.1073/pnas.211566398.

[59] G. J. McLachlan, *Discriminant analysis and statistical pattern recognition*. John Wiley & Sons, 2005.

[60] N. Abramson, D. Braverman, and G. Sebestyen, "Pattern recognition and machine learning," *IEEE Trans. Inf. Theory*, vol. 9, no. 4, pp. 257–261, 1963, doi:

10.1109/TIT.1963.1057854.
[61] P. J. Bickel and E. Levina, "Some theory for Fisher's linear discriminant function, 'naive Bayes', and some alternatives when there are many more variables than observations," *Bernoulli*, vol. 10, no. 6, pp. 989–1010, 2004, doi: 10.3150/bj/1106314847.
[62] H. Pang, T. Tong, and H. Zhao, "Shrinkage-based diagonal discriminant analysis and its applications in high-dimensional data," *Biometrics*, vol. 65, no. 4, pp. 1021–1029, 2009, doi: 10.1111/j.1541-0420.2009.01200.x.
[63] Z. Qiao, L. Zhou, and J. Z. Huang, "Sparse linear discriminant analysis with applications to high dimensional low sample size data," *IAENG Int. J. Appl. Math.*, vol. 39, no. 1, 2009.
[64] R. Moghaddass and M. J. Zuo, "Multistate degradation and supervised estimation methods for a condition-monitored device," *IIE Trans. (Institute Ind. Eng.*, vol. 46, no. 2, pp. 131–148, 2014, doi: 10.1080/0740817X.2013.770188.
[65] R. Moghaddass and M. J. Zuo, "An integrated framework for online diagnostic and prognostic health monitoring using a multistate deterioration process," *Reliab. Eng. Syst. Saf.*, vol. 124, pp. 92–104, 2014, doi: 10.1016/j.ress.2013.11.006.
[66] J. Xu, Y. Wang, and L. Xu, "PHM-oriented integrated fusion prognostics for aircraft engines based on sensor data," *IEEE Sens. J.*, vol. 14, no. 4, pp. 1124–1132, 2014, doi: 10.1109/JSEN.2013.2293517.
[67] F. O. Heimes, "Recurrent neural networks for remaining useful life estimation," *2008 Int. Conf. Progn. Heal. Manag. PHM 2008*, 2008, doi: 10.1109/PHM.2008.4711422.
[68] A. Malhi, R. Yan, and R. X. Gao, "Prognosis of defect propagation based on recurrent neural networks," *IEEE Trans. Instrum. Meas.*, vol. 60, no. 3, pp. 703–711, 2011, doi: 10.1109/TIM.2010.2078296.
[69] A. Mosallam, K. Medjaher, and N. Zerhouni, "Bayesian approach for remaining useful life prediction," *Chem. Eng. Trans.*, vol. 33, no. September, pp. 139–144, 2013, doi: 10.3303/CET1333024.
[70] A. Savitzky and M. J. E. Golay, "Smoothing and differentiation of data by simplified least squares procedures.," *Anal. Chem.*, vol. 36, no. 8, pp. 1627–1639, 1964.
[71] J. Ben Ali, B. Chebel-Morello, L. Saidi, S. Malinowski, and F. Fnaiech, "Accurate bearing remaining useful life prediction based on Weibull distribution and artificial neural network," *Mech. Syst. Signal Process.*, vol. 56, pp. 150–172, 2015, doi: 10.1016/j.ymssp.2014.10.014.
[72] L. Guo, N. Li, F. Jia, Y. Lei, and J. Lin, "A recurrent neural network based health indicator for remaining useful life prediction of bearings," *Neurocomputing*, vol. 240, pp. 98–109, 2017, doi: 10.1016/j.neucom.2017.02.045.
[73] M. Baptista, H. Prendinger, and E. Henriques, "Prognostics in Aeronautics with Deep Recurrent Neural Networks," *Phme 2020*, no. July, pp. 1–11, 2020.